\documentclass[onecolumn,11pt,aps,prd,superscriptaddress,nofootinbib,longbibliography]{revtex4-2}
\usepackage{amsmath,amssymb}
\usepackage[colorlinks,linkcolor=red,anchorcolor=blue,citecolor=cyan]{hyperref} 
\usepackage{graphicx}
\usepackage{adjustbox}
\usepackage{tikz}
\usetikzlibrary{shadings}
\usetikzlibrary{arrows.meta}
\usetikzlibrary{calc}
\usepackage{colortbl}
\usepackage{xcolor}
\usepackage{picinpar}
\usepackage{float}
\usepackage{wrapfig}

\usepackage{slashed}
\usepackage{ulem}
\usepackage{bm}
\usepackage{mathrsfs}
\usepackage{array,cases}
\usepackage{comment}
\usepackage{multirow,paracol}
\usepackage{booktabs}

\usepackage[final]{changes}
\usepackage{lipsum}
\newcommand{\diff}{\mathrm{d}}

\allowdisplaybreaks
\begin{document}
	\author{Zhiming Shuai}
	\email{202410188403@mail.scut.edu.cn}
	\affiliation{School of Physics and Optoelectronics, South China University of Technology, Guangzhou 510641, China}
	
	\author{Xiangdong Zhang}
	\email[Corresponding author: ]{scxdzhang@scut.edu.cn}
	\affiliation{School of Physics and Optoelectronics, South China University of Technology, Guangzhou 510641, China}
	
	\title{Non-singular Inflation-Dark Energy Unification Model Based on Loop Quantum Cosmology and Mass-Varying Neutrinos}
	\begin{abstract}
		Unifying the early-universe inflationary paradigm with late-time cosmic acceleration, while resolving the initial Big Bang singularity, remains one of the most profound challenges in modern cosmology. In this paper, we propose a non-singular quintessential inflation model embedded within the effective dynamics of Loop Quantum Cosmology (LQC) based on a Generalized Regularization Scheme. The quantum geometry effects naturally replace the initial singularity with a quantum bounce, followed by a phase of superinflation that sets robust initial conditions for the subsequent slow-roll inflation. To achieve a viable late-time dark energy epoch and address the coincidence problem, we introduce a coupling between the scalar field and massive neutrinos, known as Mass-Varying Neutrinos (MaVaNs). As neutrinos become non-relativistic in the post-inflationary evolution, their backreaction effectively freezes the scalar field, triggering the late-time accelerated expansion. We numerically trace the full background dynamics from the quantum bounce to the present day. Furthermore, we tightly constrain the model parameters utilizing the observational data, including the Type Ia supernovae sample, the Dark Energy Spectroscopic Instrument (DESI) Baryon Acoustic Oscillations (BAO) and Cosmic Microwave Background (CMB) distance priors. Our results demonstrate that this unified LQC-MaVaNs quintessential framework is highly consistent with current precision cosmological observations.
	\end{abstract}
	\maketitle

	\newpage
	\section{Introduction}
	One of the most profound challenges in modern cosmology is to unify the early-universe inflationary paradigm with late-time cosmic acceleration within a single theoretical framework, while simultaneously resolving the initial Big Bang singularity. Traditional cosmological models typically invoke two entirely distinct mechanisms to explain these two extreme epochs of cosmic evolution. To achieve a deeper level of theoretical unification, the quintessential inflation model\cite{Peebles_1999,de_Haro_2021,Dimopoulos_2002} was proposed. This framework utilizes a single scalar field to drive inflation at early, high-energy scales and subsequently provide dynamic dark energy behavior to account for the late-time accelerated expansion. However, within the framework of classical General Relativity, the quintessential inflation model still fails to avoid the initial Big Bang singularity.
	
	To thoroughly resolve the singularity problem, Loop Quantum Cosmology (LQC) offers a highly promising effective theory\cite{ashtekar2003mathematicalstructureloopquantum,Ashtekar_2006,Ashtekar_2006_2,Bojowald_2001,Ashtekar_2010,Singh_2006}. In LQC, quantum geometric effects naturally replace the initial singularity with a non-singular quantum bounce. Building upon this, this paper proposes closely integrating the effective dynamics of LQC with the quintessential inflation model to construct a completely non-singular, unified cosmological evolution model. Immediately following the quantum bounce, the universe naturally undergoes a phase of superinflation\cite{Singh_2006,Bojowald_2002,Copeland_2008,Ashtekar_2010,Ashtekar_2011}. This phase not only rapidly dissipates the kinetic energy of the scalar field but also establishes extremely robust initial conditions for the subsequent classical slow-roll inflation. Furthermore, to achieve a viable dark energy-dominated epoch at late times and alleviate the "coincidence problem," we introduce a coupling mechanism between the scalar field and Mass-Varying Neutrinos (MaVaNs)\cite{Fardon_2004,Brookfield_2006,Wetterich_2007,Amendola_2008,Afshordi_2005,Mota_2008}. As the universe cools during the post-inflationary evolution, neutrinos transition into non-relativistic particles. The resulting backreaction effectively freezes the evolution of the scalar field, thereby naturally triggering the late-time accelerated expansion.
	
	Another core focus of this paper is to explore and test the physical effects of different quantization schemes within the LQC framework. To systematically investigate how these choices impact cosmic evolution, we adopt a generalized effective Hamiltonian that incorporates a free regularization parameter, $\lambda$\cite{Zhang_2021}. The quantization method of LQC relies on a regularization freedom of the Hamiltonian constraint, which can be expressed through various combinations of its Euclidean and Lorentzian terms. While these combinations are completely equivalent at the classical level yielding the standard Friedmann equations, their quantization leads to fundamentally non-equivalent quantum theories. For instance, standard LQC utilizes only the Euclidean term ($\lambda = -1/\gamma^2$), whereas an alternative model treats both terms independently and equally ($\lambda = 1$). Serving as a theoretical bridge between these distinct quantization schemes, this $\lambda$-parameterized Hamiltonian provides an excellent opportunity to leverage astronomical observations to determine the optimal value of $\lambda$, thereby singling out the observationally preferred quantum theoretical foundation.
	
	In this paper, we trace the full background dynamical evolution from the quantum bounce to the present day, and rigorously confront this generalized LQC-MaVaNs model with current high-precision cosmological observational data, including the Type Ia supernovae sample\cite{Brout_2022}, Baryon Acoustic Oscillations (BAO) data from the Dark Energy Spectroscopic Instrument (DESI)\cite{Abdul_Karim_2025}, and Cosmic Microwave Background (CMB) distance priors\cite{Chen_2019}. Through this approach, we aim to explore and ultimately determine which loop quantization scheme aligns more perfectly with modern cosmological observations.
	
	The paper is organized as follows. Section \ref{sec2} reviews the effective Friedmann equations in generalized LQC. Section \ref{sec3} explores the early-universe dynamics, from the quantum bounce to slow-roll inflation. Section \ref{sec4} analyzes the post-inflationary evolution, detailing the scaling regime and the MaVaNs-triggered late-time acceleration. Section \ref{sec5} constrains the model parameters via Bayesian analysis using cosmological datasets.
	
	\section{Friedmann equation in LQC}\label{sec2}
	The Loop Quantum Cosmology (LQC) quantization method detailed in the article\cite{Zhang_2021} is based on a one-parameter regularization freedom of the Hamiltonian constraint. This approach generalizes standard LQC models and provides a framework where a small cosmological constant can emerge purely from quantum geometric effects. By introducing a free real parameter $\lambda$, a generalized expression of the classical Hamiltonian constraint is constructed as\cite{Zhang_2021}
	\begin{equation}
		H_g = \lambda H^E - (1 + \lambda \gamma^2) H^L - (1 - \lambda) \int_\Sigma\diff^3 x \sqrt{q} {}^{(3)}R,
	\end{equation}
	where $H^E$ and $H^L$ denote the Euclidean and Lorentzian terms, respectively, $\gamma$ is the Barbero-Immirzi parameter, and ${}^{(3)}R$ represents the 3-dimensional spatial curvature, which vanishes in the spatially flat Friedmann-Lemaître-Robertson-Walker (FLRW) metric. The so-called Euclidean term $H^E$ and Lorentzian term $H^L$ are denoted respectively as\cite{Ashtekar_2006_2}
	\begin{equation}
		H^E = \frac{1}{16\pi G}\int _\Sigma \diff ^3 x N F^j_{ab}\frac{\epsilon_{jkl}E^a_kE^b_l}{\sqrt{q}},
	\end{equation}
	and
	\begin{equation}
		H^L = \frac{1}{16\pi G}\int_\Sigma\diff^3x N\epsilon_{jmn}K^m_aK^n_b\frac{\epsilon_{jkl}E^a_kE^b_l}{\sqrt{q}},
	\end{equation}
	where N is the lapse function, q denotes the determinant of the spatial metric,
	$ F^i_{ab} $ is the curvature of connection $A^i_a$, and $k^i_a$ represents the extrinsic curvature of $\Sigma$. The classical phase space consists of the Ashtekar-Barbero variables ($A^i_a(x)$, $E^a_i$), where $A^i_a(x)$  is a SU(2) connection and  $E^a_i$ is a densitized triad\cite{HAN_2007,Ashtekar_2004,ashtekar2003mathematicalstructureloopquantum}. The non-vanishing Poisson bracket is given by
	\begin{equation}
		\{A^i_a(x),E^b_j(y)\} = 8\pi G \gamma \delta^b_a \delta^i_j \delta^3(x,y)
	\end{equation}
	
	Classically, any choice of $\lambda$ yields equivalent theories, specifically the standard Friedmann equations without a cosmological constant. However, the quantization of these classically equivalent expressions leads to fundamentally non-equivalent quantum theories. For instance, setting $\lambda = -1/\gamma^2$ recovers the standard LQC model utilizing only the Euclidean term\cite{Ashtekar_2006_2}, whereas $\lambda = 1$ yields an alternative model that treats the Euclidean and Lorentzian terms independently\cite{Yang_2009}. Therefore, different choices of $\lambda$ might correspond to different quantum theories. This is the case for the LQC model which we are going to consider. Our idea is to use observations to single out the preferred expression of the Hamiltonian (or the free parameter $\lambda$ ).

	Following the rigorous quantization procedure of loop quantum gravity (LQG), both the Euclidean term $H^E$ and the Lorentzian term $H^L$ are promoted to well-defined difference operators acting on the kinematical Hilbert space\cite{Zhang_2021}. To investigate the physical implications, the effective Hamiltonian constraint can be derived at the leading order as\cite{Zhang_2021}
	\begin{equation}\label{effective hamiltonian constarint}
		H_F = -\frac{3\beta}{8\pi G\gamma^2\Delta} |v| \sin^2(b) \left[1 - (1 + \lambda\gamma^2)\sin^2(b)\right] + \beta|v|\rho \approx 0,
	\end{equation}
	where $v$ and $b$ are the canonical conjugate variables characterizing the quantum geometry, $\Delta = 4\sqrt{3}\pi\gamma G \hbar$ is the minimum non-zero area eigenvalue, $\beta = 2\pi G\hbar\gamma\sqrt{\Delta}$, and $\rho$ is the total energy density of the matter field. The effective equations of motion of the model with respect to the cosmological time t can be derived by the Hamiltonian constraint \eqref{effective hamiltonian constarint}. expecially, the the Friedmann equation are derived as\cite{Zhang_2021}
	\begin{equation}\label{Friedmann equation}
		H^2 = \left(\frac{\dot{v}}{3v}\right)^2 = \frac{1}{\gamma^2\Delta}\sin^2(b)(1-\sin^2(b))(1-2(1+\lambda\gamma^2)\sin^2(b))^2.
	\end{equation}
	
	By utilizing the vanishing of the effective Hamiltonian constraint ($H_F \approx 0$), one can solve for the geometric quantity $\sin^2(b)$ as a function of the matter energy density $\rho$, namely
	\begin{equation}\label{sin2b}
		\sin^2(b_\pm)=\frac{1\pm\sqrt{1-\frac{\rho}{\rho_c}}}{2(1+\lambda\gamma^2)}.
	\end{equation}
	where $\rho_c$ is the critical energy density, denoting the maximum allowed density in this generalized LQC model, given by
	\begin{equation}
		\rho_c = \frac{3}{32\pi G \gamma^2 \Delta (1+\lambda\gamma^2)}.
	\end{equation}
	Substituting Eq. \eqref{sin2b} back into Eq.\eqref{Friedmann equation} allows us to express the Hubble parameter entirely in terms of the energy density
	\begin{equation}\label{Friedmann_equation_rho}
		H^2 = \frac{1}{4\Delta\gamma^2(1+\lambda\gamma^2)^2}\left( 1-\frac{\rho}{\rho_c} \right)\left( 1\pm\sqrt{1-\frac{\rho}{\rho_c}} \right)\left(1+2\lambda\gamma^2\mp \sqrt{1-\frac{\rho}{\rho_c}}\right),
	\end{equation}
	The presence of the $\pm$ branches in the modified Friedmann equation \eqref{Friedmann_equation_rho} originates from the quadratic nature of the constraint equation with respect to $\sin^2(b)$. Specifically, the positive branch ($+$) dictates that in the low-energy limit where the matter energy density is significantly lower than the critical density ($\rho \ll \rho_c$), the quantum geometric corrections naturally decouple, elegantly recovering the standard Friedmann equation of general relativity. Conversely, the negative branch ($-$) corresponds to an alternative cosmological phase. In the same low-energy limit, rather than reducing to standard general relativity, it leads to a modified Friedmann equation characterized only by an emergent effective cosmological constant and a slightly rescaled effective Newtonian constant.
	

	\section{Superinflation and Inflation}\label{sec3}
	In this section, we shall investigate the superinflation and inflation within the framework of the effective theory we have introduced. Within the LQC framework, the classical Big Bang singularity is replaced by a non-singular quantum bounce that smoothly connects the contracting and expanding phases of the universe. Naturally, this bounce establishes the initial conditions for cosmic inflation. To address the evolutionary history from the quantum bounce to the classical slow-roll regime, we specify the background dynamics driven by a scalar field $\phi$ endowed with a generalized exponential potential. Specifically, the potential is parameterized as\cite{Geng_2015,Geng_2017}
	\begin{equation}\label{potenrtial}
		V(\phi) = V_0 \exp\left[-\alpha \left(\frac{\phi}{m_{\text{Pl}}}\right)^n\right],
	\end{equation}
	where $V_0$ represents the energy scale of the potential, $\alpha$ and $n$ are dimensionless free parameters shaping the potential curve and $m_{Pl}=1/\sqrt{G}$ is the Planck mass in natural units ($\hbar = c = 1$). For the purpose of our numerical and Bayesian analyses in this paper, we fix the exponent to $n=6$.
	
	The physical motivation for adopting this specific functional form of the potential is manifold. First, it serves as an excellent candidate for the quintessential inflation paradigm. Standard cosmological models typically invoke two distinct mechanisms to explain the early-universe inflation and the late-time cosmic acceleration. This generalized potential elegantly unifies these two extreme epochs, namely its flat plateau at high field values drives the primordial inflation, while its asymptotic tail provides a dynamic dark energy behavior capable of confronting late-time observational constraints. Second, the integration of this potential into the LQC effective dynamics is physically robust. Since the required inflationary energy scale $V_0$ is significantly lower than the Planckian critical density ($V_0 \ll \rho_c$), a potential-dominated quantum bounce is strictly prohibited. Consequently, the universe must bounce in a state dominated by immense kinetic energy ($\dot{\phi}^2/2 \approx \rho_c$). Following the bounce, the scalar field undergoes a characteristic LQC superinflation phase ($\dot{H} > 0$), which rapidly dissipates the kinetic energy via intense Hubble damping. This unique dynamical attractor mechanism naturally decelerates and freezes the scalar field on the flat region of $V(\phi)$, perfectly setting the initial conditions for the subsequent slow-roll inflation. Finally, by employing this specific potential, the model can seamlessly extend into the post-inflationary evolution without invoking any additional dark energy fields or a cosmological constant. Because this single quintessential field dictates the entire cosmic history, it establishes a continuous and unbroken theoretical link between the primordial dynamics and the present day. This robust tracking behavior allows us to extrapolate the early-universe framework into late-time cosmic expansion, ultimately making it possible to rigorously test and constrain the unified model using current cosmological datasets.
	
	While the phenomenon of superinflation within the basic LQC framework has been extensively investigated in the literature \cite{Copeland_2008,Bojowald_2011,Ashtekar_2010,Singh_2006,Bojowald_2002,Ashtekar_2011,Agullo_2012,Artymowski_2009,Barboza_2022,Bhardwaj_2019,Bonga_2016,Zhang_2007,Bonga_2016_2}, we provide a brief dynamical analysis here to ensure the present paper remains self-contained. More importantly, our specific focus in this section is to elucidate the novel physical implications introduced by the generalized regularization parameter $\lambda$ during this extreme early-universe epoch. To this end, we begin with the equation of motion for the scalar field, described by the Klein-Gordon equation
	\begin{equation}
		\ddot{\phi}+3H\dot{\phi}+\frac{\mathrm{d}V(\phi)}{\mathrm{d}\phi}=0.
	\end{equation}
	Here, the dependence on $\lambda$ is implicit in the Hubble parameter $H$.
	
	However, during the superinflationary epoch, the kinetic energy of the scalar field overwhelmingly dominates the background dynamics ($\dot{\phi}^2/2 \gg V(\phi)$). Consequently, the cosmic evolution is largely insensitive to the specific shape of the quintessential potential. Under this kinetic-dominated approximation, the potential gradient term can be safely neglected, reducing the Klein-Gordon equation to a much simpler form:
	\begin{equation}
		\ddot{\phi}+3H\dot{\phi}=0.
	\end{equation}
	By integrating this simplified equation in conjunction with the generalized Friedmann equation Eq.~\eqref{Friedmann_equation_rho}, we can analytically solve for the trajectory of the scalar field
	\begin{equation}
		\frac{\mathrm{d}\phi}{\mathrm{d}N} = \left\{ \begin{aligned}
			\frac{C_1 \mathrm{e}^{-\mathrm{ArcTanh}\left( \frac{1+\lambda\gamma^2+\sqrt{1-e^{-6N}}}{\lambda\gamma^2}\right)}}{\sqrt{1-\mathrm{e}^{-6N}}}, \qquad & \text{positive branch} \\
			\frac{C_2 \mathrm{e}^{-\mathrm{ArcTanh}\left( \frac{1+\lambda\gamma^2-\sqrt{1-e^{-6N}}}{\lambda\gamma^2}\right)}}{\sqrt{1-\mathrm{e}^{-6N}}}, \qquad & \text{negative branch} 
		\end{aligned}\right.
	\end{equation}
	and
	\begin{equation}
		\phi(N)= \left\{ \begin{aligned}
			&\phi_0 + C_2\frac{N}{\sqrt{1+\lambda\gamma^2}} -C_2\frac{4+3\lambda\gamma^2}{48(1+\lambda\gamma^2)^{3/2}}\mathrm{e}^{-6N} + \mathcal{O}(\mathrm{e}^{-12N}), \quad & \text{positive branch} \\&
			\phi_0 + C_2\frac{1}{6\sqrt{\lambda\gamma^2}}\mathrm{e}^{-3N}+C_2\frac{(1-5\lambda\gamma^2)}{144\lambda^{3/2}\gamma^3} + \mathcal{O}(e^{-15N}), \quad & \text{negative branch} 
		\end{aligned}\right.
	\end{equation}
	where $C_i$ is the integration coefficient. Meanwhile, the explicit dependence of Hubble Parameter on the scale factor during this epoch can be analytical obtained via $\rho=\rho_c/a^6$. Therefore, the Friedmann equation cast into the form
	\begin{equation}
		H^2 = \frac{1}{4\Delta\gamma^2(1+\lambda\gamma^2)^2}\left( 1-a^{-6} \right)\left( 1\pm\sqrt{1-a^{-6}} \right)(1+2\lambda\gamma^2\mp \sqrt{1-a^{-6}}).
	\end{equation}
	This explicit analytical expression reveals how the generalized regularization parameter $\lambda$ fundamentally dictates the background expansion. The presence of $\lambda$ not only suppresses the overall amplitude of the Hubble parameter via the global prefactor but also dynamically reshapes the evolutionary trajectory through the bracketed terms. Consequently, a non-zero $\lambda$ directly modulates the maximum expansion rate ($H_{\text{max}}$) during superinflation and the corresponding intensity of the Hubble friction, offering a theoretical mechanism to control how the universe transitions into the subsequent slow-roll phase.

	\begin{figure}[htbp]
		\centering
		\begin{minipage}{0.3\linewidth}
			\centering
			\includegraphics[width=\linewidth]{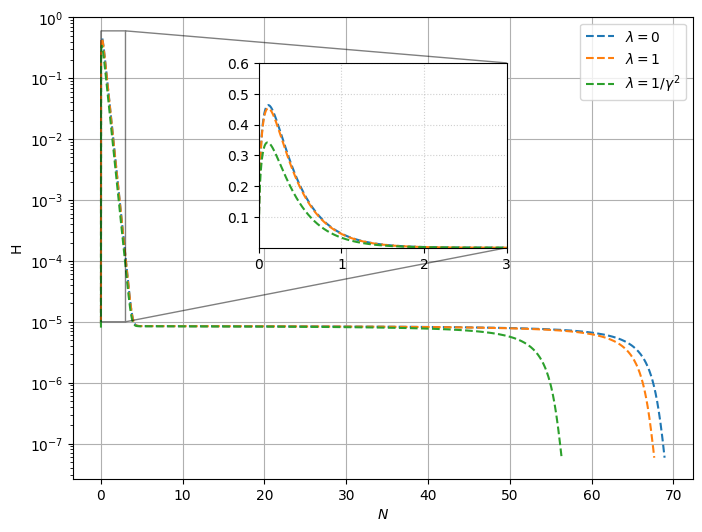}\\[0.1cm]
			(a)
		\end{minipage}
		\hfill
		\begin{minipage}{0.3\linewidth}
			\centering
			\includegraphics[width=\linewidth]{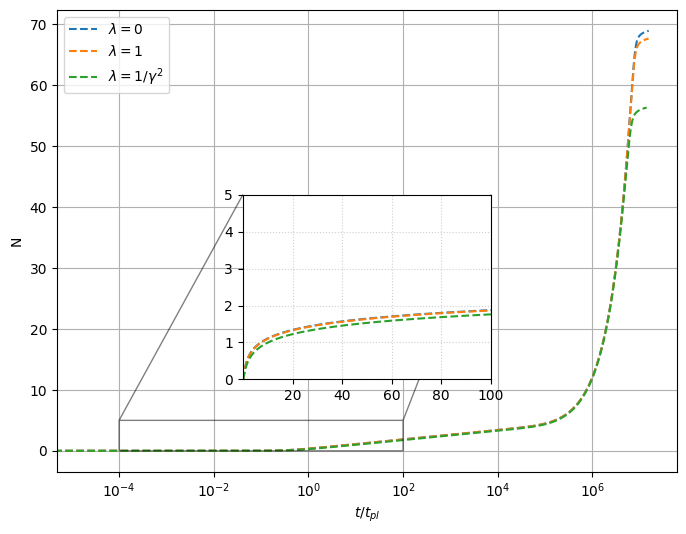}\\[0.1cm]
			(b)
		\end{minipage}
		\hfill
		\begin{minipage}{0.3\linewidth}
			\centering
			\includegraphics[width=\linewidth]{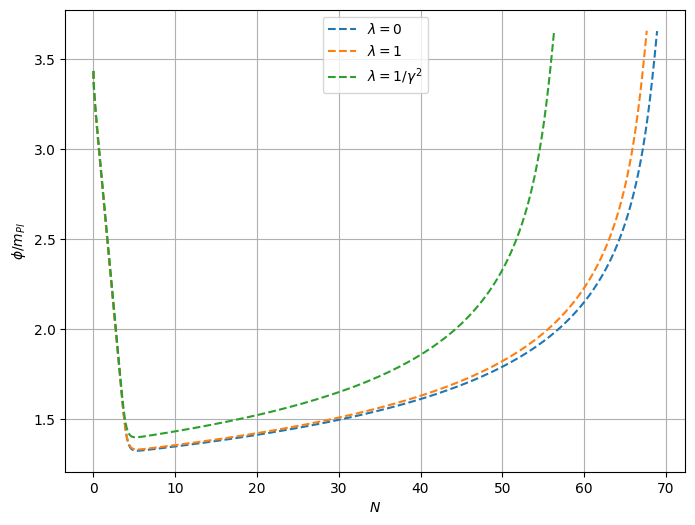}\\[0.1cm]
			(c)
		\end{minipage}
		\caption{The background cosmological evolution following the quantum bounce for different values of the Lorentzian parameter $\lambda$. (\textbf{a}) The evolution of the Hubble parameter $H$ with respect to the number of e-folds $N$. The inset zooms in on the extremely early stage ($N \in [0,3]$), illustrating the superinflation phase ($\dot{H}>0$). (\textbf{b}) The accumulation of e-folds $N$ as a function of the cosmic time. The horizontal plateaus in $H$ and the subsequent steep rises in $N$ represent the classical slow-roll inflationary phase. (\textbf{c}) The trajectory of the scalar field $\phi$ with respect to the number of e-folds $N$.  In the numerical integrations presented here, the dynamics are explicitly computed for the branch without the emergent cosmological constant. The potential parameters are set to $V_0 = 8.874\times10^{-12} m_{\text{Pl}}^4$, $\alpha = 5.426\times10^{-3}$, and $n=6$. The initial field value at the bounce is chosen as $\phi_B = 3.432 m_{\text{Pl}}$, with the Barbero-Immirzi parameter $\gamma = 0.2375$. 
		}
		\label{fig:lna_H}
	\end{figure}
	
	To corroborate the analytical derivations and gain deeper insight into the cosmological dynamics, we numerically integrate the background equations of motion. The results are presented in Fig.~\ref{fig:lna_H}, which illustrates the evolutionary history of the universe from the quantum bounce through the slow-roll inflationary phase for distinct values of the generalized LQC parameter $\lambda$. It is worth noting that, we only numerically compute the positive branch, for the alternative branch with an effective cosmological constant, its evolutionary trajectories are practically indistinguishable from the $\lambda=0$ case owing to the extremely small magnitude of $\lambda$. In Fig.~\ref{fig:lna_H}(a), we depict the evolution of the Hubble parameter $H$ against the number of e-folds $N$. Immediately following the bounce at $N=0$, the universe enters the superinflationary phase characterized by a rapidly increasing expansion rate ($\dot{H} > 0$). As analytically anticipated, the peak amplitude of the Hubble parameter, $H_{\text{max}}$, is highly sensitive to the parameter $\lambda$. Specifically, increasing $\lambda$ from $0$ (the standard LQC limit) to $1/\gamma^2$ markedly suppresses the peak value of $H$. Subsequently, $H$ decays rapidly before stabilizing into a constant plateau, marking the onset of classical slow-roll inflation. This transition is directly reflected in the scalar field trajectory in Fig.~\ref{fig:lna_H}(c). Initially kinetic-dominated, $\phi$ drops rapidly from its initial value. The intense Hubble friction during superinflation dissipates this kinetic energy, causing the field to "freeze" and turn around. Crucially, a larger $\lambda$ generates less Hubble friction (due to a lower $H_{\text{max}}$), causing the field to turn around earlier and at a slightly higher minimum value before potential energy dominates and slow-roll begins. Fig.~\ref{fig:lna_H}(b). Finally, Fig.~\ref{fig:lna_H}(b) illustrates the accumulation of e-folds $N$ over cosmic time. It clearly demonstrates that $\lambda$ dictates the total duration of inflation. While the standard LQC scenario ($\lambda=0$) yields $N \approx 69$ total e-folds, introducing a non-zero $\lambda$ alters the initial kinetic phase and the precise onset of slow-roll, observably reducing the total inflationary duration (e.g., $N \approx 56$ for $\lambda=1/\gamma^2$).
	
	Following the intense Hubble damping during the superinflation phase, the scalar field loses its kinetic energy and gently lands on the flat plateau of the  potential as described in Fig.~\ref{fig:lna_H}(c). At this stage, the universe enters the classical slow-roll inflationary regime where the potential energy dominates over the kinetic energy ($V(\phi) \gg \dot{\phi}^2/2$).  To quantitatively describe the flatness of the potential and the duration of this phase, it is customary to introduce the slow-roll parameters
	\begin{equation}\label{slow roll parameters}
		\epsilon_V \equiv \frac{M_{\text{Pl}}^2}{2}\left(\frac{V_{,\phi}}{V}\right)^2, \quad \quad \eta_V \equiv M_{\text{Pl}}^2 \frac{V_{,\phi\phi}}{V}.
	\end{equation}
	The condition for sustained inflation translates to $\epsilon_V \ll 1$ and $|\eta_V| \ll 1$. Inflation naturally terminates when the field rolls into the steeper region of the potential, breaking the slow-roll condition, typically marked by $\epsilon_V \simeq 1$.
	
	During this quasi-exponential expansion, the scalar spectral index $n_s$ and tensor‑to‑scalar ratio $r$ are the key observables connecting these primordial perturbations to CMB anisotropies. Although the very early universe is governed by the LQC effective dynamics, the background energy density has plummeted several orders of magnitude below the Planckian critical density ($\rho \ll \rho_c$) in the inflationary epoch. Consequently, the spacetime geometry effectively reduces to classical general relativity. Therefore, it is entirely mathematically rigorous and physically justified to employ the standard classical slow-roll approximation to estimate both the scalar spectral index and the tensor-to-scalar ratio\cite{Zhang_2007}
	\begin{align}
		n_s &\simeq 1 - 6\epsilon_{V} + 2\eta_{V}, \label{eq:ns_approx} \\
		r &\simeq 16\epsilon_{V}. \label{eq:r_approx}
	\end{align}
	
	\begin{figure}[htbp]
		\centering
		\begin{minipage}{0.3\linewidth}
			\centering
			\includegraphics[width=\linewidth]{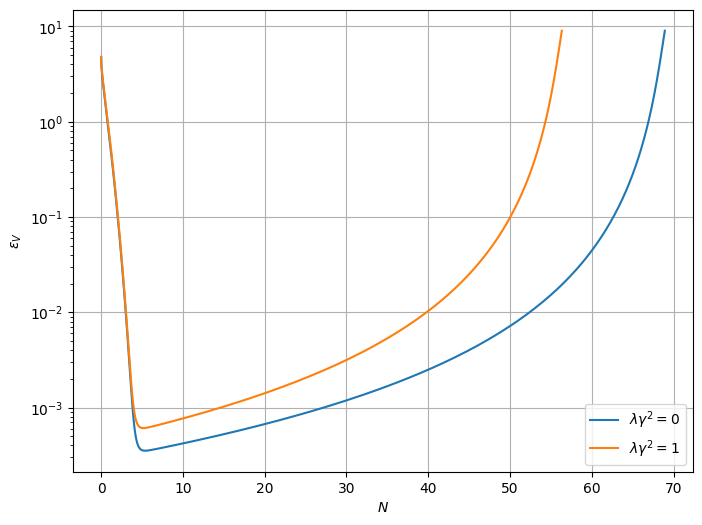}\\[0.1cm]
			(a)
		\end{minipage}
		\vspace{0.5cm}
		\begin{minipage}{0.3\linewidth}
			\centering
			\includegraphics[width=\linewidth]{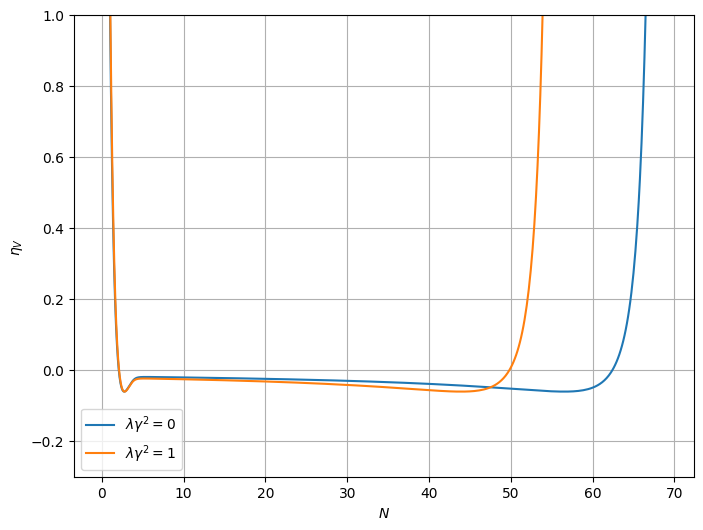}\\[0.1cm]
			(b)
		\end{minipage}
		\vspace{0.5cm}
		\begin{minipage}{0.3\linewidth}
			\centering
			\includegraphics[width=\linewidth]{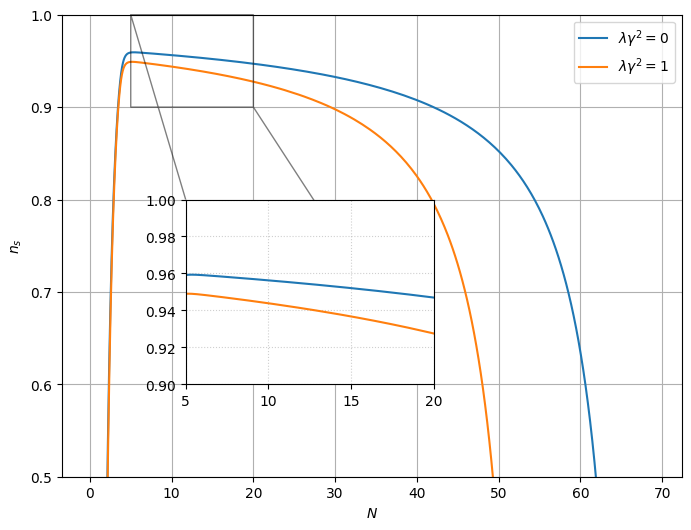}\\[0.1cm]
			(c)
		\end{minipage}
		
		\caption{Evolution of the slow-roll parameters and the scalar spectral index during the inflationary phase. (\textbf{a}) The first slow-roll parameter $\epsilon_V$. (\textbf{b}) The second slow-roll parameter $\eta_V$. (\textbf{c}) The scalar spectral index $n_s$. The inset highlights the observable window ($N \in [5, 20]$).}
		\label{fig:slow_roll_params}
	\end{figure}
	
	To further quantitatively analyze the slow-roll inflationary dynamics and observables, Fig.~\ref{fig:slow_roll_params} presents the evolution of the slow-roll parameters and the scalar spectral index. As shown in Figs.~\ref{fig:slow_roll_params}(a) and \ref{fig:slow_roll_params}(b), following the intense Hubble damping during the superinflation phase, the slow-roll parameters $\epsilon_V$ and $\eta_V$ rapidly drop well below unity, confirming the robust onset of the slow-roll regime. Inflation ultimately terminates when the slow-roll condition is violated at $\epsilon_V \simeq 1$. The generalized LQC parameter $\lambda$ significantly impacts both the evolution of these parameters and the duration of inflation. Introducing a non-zero $\lambda$ ($\lambda\gamma^2 = 1$) alters the field dynamics, visibly elevating $\epsilon_V$ across the observable window and causing it to reach unity earlier than in the standard limit ($\lambda=0$).  Furthermore, because the tensor-to-scalar ratio $r$ is directly determined by the first slow-roll parameter\eqref{slow roll parameters}, this elevation of $\epsilon_V$ under $\lambda\gamma^2 = 1$ predicts a relatively larger $r$. Fig.~\ref{fig:slow_roll_params}(c) depicts the corresponding evolution of the scalar spectral index $n_s$. In the standard LQC limit ($\lambda=0$), immediately following the kinetic-dominated superinflation phase, $n_s$ rapidly climbs to a highly stable plateau. For the pivot scale $k_* = 0.05 \, \text{Mpc}^{-1}$ (corresponding to horizon exit roughly 60 e-folds before the end of inflation, around $N \in [5, 10]$ in our numerical timeline\cite{planck2020,Liddle_2003,Dimopoulos_2017}), the model predicts $n_s \approx 0.96$. As highlighted in the inset of Fig.~\ref{fig:slow_roll_params}(c), This prediction is close to the Planck 2018 result, in which $n_s = 0.9649$\cite{planck2020}. Crucially, however, introducing a non-zero $\lambda$ ($\lambda\gamma^2 = 1$) systematically suppresses the amplitude of $n_s$ across the entire observable window. This robustly demonstrates that generalized quantum geometric modifications at the deep Planck scale not only reshape the very early background universe but also cascade forward to leave distinct microscopic imprints on the observable primordial power spectrum.

	\section{Post-Inflation Evolution}\label{sec4}
	After the end of the slow-roll inflation, we now turn our attention to the post-inflationary history of the universe. In quintessential inflation, the scalar field must span the radiation and matter-dominated eras before eventually re-emerging to drive the late-time cosmic acceleration. Unlike standard inflation in which the scalar field oscillates around a potential minimum to reheat the universe, the scalar field continues to roll down its potential, exhibiting a runaway behavior. To successfully transition to the thermalized Hot Big Bang epoch,  particle production must take place via instant particle production\cite{Felder_1999} or gravitational particle creation\cite{PhysRevD.35.2955,PhysRevD.42.453}. 
	
	After the end of inflation, the kinetic energy of the scalar field overwhelmingly dominates over its potential energy ($\dot{\phi}^2/2 \gg V(\phi)$). The universe thereby enters a unique evolutionary epoch known as the kination phase\cite{}. In this kinetic-dominated regime, the effective equation of state of the scalar field is $w_\phi \simeq 1$. A direct consequence of this equation of state is that the energy density of the quintessential field dilutes extremely rapidly with the expansion of the universe, scaling as $\rho_\phi \propto a^{-6}$. Therefore, the kinetic energy of the scalar field decreases faster than that of radiation, after which radiation dominates the cosmic evolution.
	
	Following the kination era, the universe seamlessly enters the standard radiation-dominated (RD) and subsequent matter-dominated (MD) epochs. During these eras, the scalar field energy density exhibits a ``scaling'' behavior, dynamically evolving in tandem with the dominant background fluid. Namely, rather than fully diluting with cosmic expansion or prematurely dominating the cosmic expansion, the scalar field converges toward a stable dynamical attractor. Along this tracking trajectory, the field's effective equation of state automatically adjusts to mimic that of the prevailing background ($w_\phi \simeq 1/3$ during the RD era, and $w_\phi \simeq 0$ during the MD era). Consequently, the scalar field energy density $\rho_\phi$ scales proportionally to the background energy density. To mathematically substantiate this dynamic tracking mechanism and strictly demonstrate the existence of the attractor, we reformulate the cosmological evolution equations into an autonomous dynamical system. We introduce the standard dimensionless variables characterizing the fractional kinetic and potential energies of the scalar field,namely
	\begin{equation}
		x \equiv \frac{\dot{\phi}}{\sqrt{6} M_{\text{Pl}} H}, \quad \quad y \equiv \frac{\sqrt{V(\phi)}}{\sqrt{3} M_{\text{Pl}} H},
	\end{equation}
	along with the background fluid density parameter $\Omega_B \equiv \rho_B / (3 M_{\text{Pl}}^2 H^2)$, where the subscript $B$ denotes the dominant background fluid. The fractional energy density and the effective equation of state of the scalar field are expressed as $\Omega_\phi = x^2 + y^2$ and $w_\phi = (x^2 - y^2)/(x^2 + y^2)$, respectively. By taking the derivatives with respect to the number of e-folds $N = \ln a$, the Klein-Gordon equations can be cast into the following formalism
	\begin{align}
		\frac{dx}{dN} &= -3x + \sqrt{\frac{3}{2}} \lambda_{eff} y^2 + \frac{3}{2} x \left( x^2+y^2+(1+w_B)(1-x^2-y^2) \right), \label{eq:dx_dN} \\
		\frac{dy}{dN} &= -\sqrt{\frac{3}{2}} \lambda_{eff} x y + \frac{3}{2} y \left( x^2 +y^2 + (1 + w_B)(1 - x^2 - y^2) \right), \label{eq:dy_dN}
	\end{align}
	where $\lambda_{eff} \equiv -M_{\text{Pl}} V_{,\phi} / V$ represents the effective slope of the quintessential potential. By setting $dx/dN = 0$ and $dy/dN = 0$, we can determine the critical points of this autonomous system. By setting $dx/dN = 0$ and $dy/dN = 0$, we can determine the critical points of this autonomous system. Assuming a physically viable regime with a subdominant field ($\Omega_\phi < 1$) and non-vanishing potential ($y \neq 0$), the dynamical system yields a well-known scaling attractor solution.
	
	To naturally exit the matter-dominated scaling regime and trigger the late-time cosmic acceleration, we introduce a direct interaction between the scalar field and massive neutrinos. We propose that the neutrino mass is a dynamical quantity governed by
	\begin{equation}
		m_\nu(\phi)=m_0\mathrm{e}^{\beta(\phi/m_{Pl})^n}.
	\end{equation}

	The energy density stored in the neutrinos is given by 
	\begin{equation}\label{neutrino energy density}
		\rho_\nu = \frac{g_s}{a^4}\int \frac{\mathrm{d}^3q}{(2\pi)^3}\sqrt{q^2+m_\nu^2a^2}f_\nu(q),
	\end{equation}
	and the pressure reads
	\begin{equation}
		p_\nu=\frac{g_s}{a^4}\int \frac{\mathrm{d}^3q}{(2\pi)^3}\frac{q^2}{3\sqrt{q^2+m_\nu^2a^2}}f_\nu(q),
	\end{equation}
	where $f_\nu(q)$ denotes the Fermi-Dirac distribution function
	\begin{equation}
		f_\nu(q)=\frac{1}{\mathrm{e}^{q/k_BT_{\nu,0}}+1},
	\end{equation}
	and $g_s$ represents the number of degrees of freedom for neutrino and $k_B$ is the Boltzmann's constant. Tanking the time derivative of Eq.~\eqref{neutrino energy density}, it can be easily shown that
	\begin{equation}
		\dot{\rho}_\nu + 3H(\rho_\nu+p_\nu)=\frac{\mathrm{d}\ln m_\nu(\phi)}{\mathrm{d}\phi}\dot{\phi}(\rho_\nu-3p_\nu).
	\end{equation}
	The dynamics of the scalar field can be inferred from the Klein-Gordon equation, now including an additional term arising from neutrino coupling, namely
	\begin{equation}\label{modified klein-gordon equation}
		\ddot{\phi}+3H\dot{\phi}+\frac{\mathrm{d}V(\phi)}{\mathrm{d}\phi}=-\frac{\mathrm{d}\ln m_\nu(\phi)}{\mathrm{d}\phi}(\rho_\nu-3p_\nu).
	\end{equation}
	From Eq.~\eqref{modified klein-gordon equation}, the pressure of neutrinos obeys the radiation‑like equation of state $p_\nu \simeq \rho_\nu/3$ as long as they remain highly relativistic. Consequently, the source term $(\rho_\nu - 3p_\nu)$ identically vanishes. Therefore, this coupling term remains entirely inactive at early times, guaranteeing that the interaction does not spoil the standard tracking behavior of the scalar field throughout the radiation and early matter-dominated eras. It is only when the universe expands and cools sufficiently that the neutrinos become non-relativistic ($p_\nu \ll \rho_\nu$), causing $\rho_\nu - 3p_\nu \simeq \rho_\nu$ to become non-zero. This transition acts as a dynamical trigger, activating the coupling term on the right-hand side of Eq.~\eqref{modified klein-gordon equation} to drag the scalar field out of its scaling regime and initiate the late-time cosmic acceleration.
	
	After introducing the coupling term with neutrinos, the evolution equations \eqref{modified klein-gordon equation} for the scalar field become
	\begin{align}
		\frac{\mathrm{d}x}{\mathrm{d}N}&= -\frac{3}{2}w_B x^3+\frac{\sqrt{6}}{2}(\lambda_{eff}y^2+(1-3w_\nu)\beta_{eff}\Omega_\nu )-\frac{3}{2}x(1-w_\nu\Omega_\nu+w_{B}(-1+y^2+\Omega_\nu)), \\ 
		\frac{\mathrm{d}y}{\mathrm{d}N}&= -\frac{\sqrt{6}}{2}\lambda_{eff}xy+\frac{3}{2}y(x^2+y^2+(1+w_B)(1-x^2-y^2-\Omega_n)+(1+w_\nu)\Omega_\nu ),
	\end{align}
	where $\beta_{eff}=-M_{Pl}\frac{\mathbf{d}\ln m_\nu(\phi)}{\mathbf{d}\phi}$. By evaluating the fixed points of this autonomous system, we obtain
	\begin{align}
		\Omega_\phi &= \frac{\beta}{\alpha}\Omega_n + \frac{3+w_{B}-3w_B\Omega_\nu}{\lambda_{eff}^2} + \mathcal{O}\left(\frac{1}{\lambda_{eff}^3}\right), \\ w_\phi &= -1 - \frac{3(1+w_B(1-\Omega_n))^2}{\lambda_{eff}\beta_{eff}\Omega_n} + O\left(\frac{1}{\lambda_{eff}^2}\right).
	\end{align}
	Therefore, this coupled term naturally ensures that the scalar field freezes at late times, yielding an equation of state that approaches the cosmological constant limit ($w_\phi \simeq -1$). Furthermore, the present-day fractional energy density of dark energy, $\Omega_\phi$, is solely determined by the ratio of the model parameters ($\beta/\alpha$). This parameter-dependent fixed point provides an elegant and straightforward resolution to the cosmic coincidence problem, as it dynamically drives the universe to the currently observed dark energy proportions without the need for fine-tuned initial conditions.
	
	\begin{figure}
		\centering
		\begin{minipage}{0.3\linewidth}
			\centering
			\includegraphics[width=\linewidth]{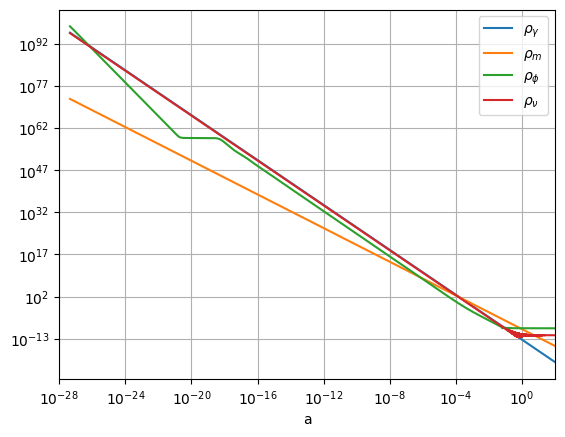}\\[0.1cm]
			(a)
		\end{minipage}
		\begin{minipage}{0.3\linewidth}
			\centering
			\includegraphics[width=\linewidth]{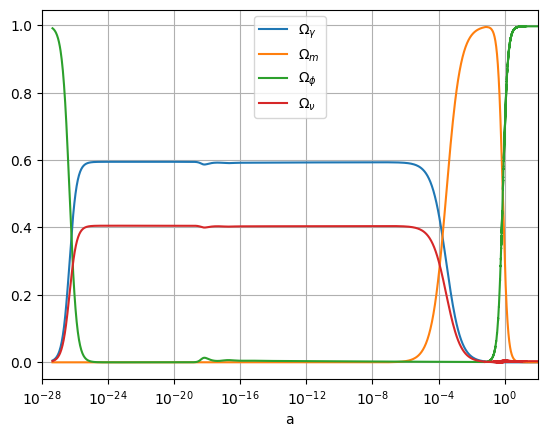}\\[0.1cm]
			(b)
		\end{minipage}
		\begin{minipage}{0.3\linewidth}
			\centering
			\includegraphics[width=\linewidth]{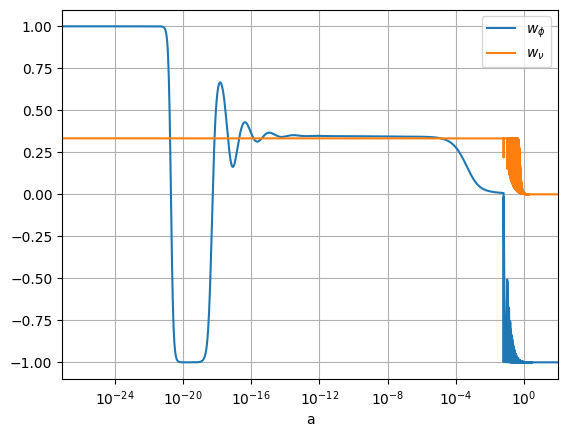}\\[0.1cm]
			(c)
		\end{minipage}
		\caption{Numerical evolution of the cosmological background from the post-inflationary epoch to the present day as a function of the scale factor $a$. (\textbf{a}) Evolution of the energy densities $\rho_i$. Following inflation, the scalar field experiences a kination phase ($\rho_\phi \propto a^{-6}$), eventually subduing to the radiation density. At late times, the coupling modifies the trajectory, forming a dark energy plateau. (\textbf{b}) The fractional energy densities $\Omega_i$. The scalar field remains strictly subdominant ($\Omega_\phi \ll 1$) during the radiation and matter eras, and gracefully emerges to dominate the universe ($\Omega_\phi \sim 0.7$) at the present epoch. (\textbf{c}) The equations of state $w_i$. The tracking behavior is clearly observed as $w_\phi$ mimics radiation ($w_\phi \simeq 1/3$) and subsequently matter ($w_\phi \simeq 0$). The dense fluctuations near $a \sim 10^{-1}$ reflect the dynamical response of the scalar field as it settles into the effective potential minimum induced by the non-relativistic transition of neutrinos ($w_\nu \to 0$).}
		\label{fig:background_evolution}
	\end{figure}
	
	To quantitatively verify the analytical predictions and confront our model with observations, we numerically integrate the exact background equations from the end of inflation to the present day. The comprehensive evolutionary history of the universe is presented in Fig.~\ref{fig:background_evolution}. The Fig.~\ref{fig:background_evolution}(a) illustrates the evolution of the energy densities. Following the initial conditions at extremely small scale factors, the scalar field initiates the kination phase, diluting sharply as $\rho_\phi \propto a^{-6}$ with an equation of state $w_\phi \simeq 1$. The radiation scales as $a^{-4}$, eventually overtaking the scalar field and marking the commencement of the standard Radiation-Dominated (RD) era. Once captured by the dynamical tracking attractor, the quintessential field's equation of state $w_\phi$ automatically adjusts itself to mimic that of the dominant background fluid, stabilizing at $w_\phi \simeq 1/3$ during the RD era and $w_\phi \simeq 0$ during the Matter-Dominated (MD) era. Consequently, the fractional energy density $\Omega_\phi$ remains suppressed well below $0.01$ throughout the early and intermediate epochs. This robust scaling behavior ensures that our model respects the stringent observational limits imposed by Big Bang Nucleosynthesis (BBN)\cite{}. The coupled mechanism takes effect in the late‑time regime ($a \gtrsim 10^{-2}$). As the universe expands, the cosmic temperature drops below the neutrino mass scale, rendering the neutrinos non-relativistic. This phase transition is captured in Fig.~\ref{fig:background_evolution}(b), where the neutrino equation of state $w_\nu$ dramatically drops from $1/3$ to $0$. This transition acts as a rigorous dynamical trigger, instantly activating the coupling source term and breaking the scaling regime. 
	
	\section{Observational Constraints}\label{sec5}
	In this section, we adopt the observational data to systematically constrain the parameter space of our model.To place rigorous constraints on the model, we utilize a combination of geometric and distance-ladder measurements. The analyzed parameter space spans the standard cosmological variables ($\Omega_m, H_0$), the coupled scalar field parameters ($\alpha, \beta, V_0$), and the LQC corrections ($\lambda, \gamma$), with a specific focus on varying the Barbero-Immirzi parameter $\gamma$ to derive its observational bounds. 
	Since our focus lies on the background evolution, we utilize the compressed Cosmic Microwave Background (CMB) distance priors\cite{Chen_2019} from the Planck 2018 mission\cite{planck2020}, including the shift parameter $R$, the acoustic scale $\ell_A$, and the baryon density $\omega_b$. These priors provide a robust high-redshift anchor at the last scattering surface. To trace the intermediate expansion history, we incorporate the latest Baryon Acoustic Oscillations (BAO) measurements from the DESI results \cite{Abdul_Karim_2025}. The precise transverse and radial distances ($D_M/r_d$ and $D_H/r_d$) are particularly sensitive to the transition into the coupled dark energy epoch. For the late-time accelerating expansion, we adopt the Pantheon+ Type Ia Supernovae (SN Ia) compilation\cite{Brout_2022}, utilizing the version calibrated with the SH0ES Cepheid-based distance ladder \cite{Riess_2022}. This inclusion incorporates local geometric measurements to map recent dynamics. To systematically explore the parameter space and derive constraints from the joint datasets of these independent cosmological probes, we employ the nested sampling algorithm \texttt{PolyChord} \cite{Handley_2015, Handley_2015_2} implemented within the Python package \texttt{cobaya} \cite{Torrado_2021}.
	
	To perform a comprehensive model comparison, we calculate the Bayesian evidence $\mathcal{Z}$, which integrates the likelihood over the entire prior volume. The model comparison is quantified by the log-Bayes factor, $\Delta \ln \mathcal{Z} = \ln \mathcal{Z}_{1} - \ln \mathcal{Z}_{2}$. A positive value indicates a statistical preference for the model. The strength of this preference is interpreted according to the revised empirical Jeffreys' scale \cite{Kass01061995, Trotta_2008}, as summarized in Table~\ref{tab:jeffreys_scale}.
	
	\begin{table}[htbp]
		\centering
		\renewcommand{\arraystretch}{1.2}
		\begin{tabular}{cl}
			\hline\hline
			$|\Delta \ln \mathcal{Z}|$ & Empirical Interpretation \\
			\hline
			$[0, 1)$ & Inconclusive (weak evidence) \\
			$[1, 3)$ & Positive (definite) evidence \\
			$[3, 5)$ & Strong evidence \\
			$\ge 5$  & Decisive evidence \\
			\hline\hline
		\end{tabular}
		\caption{The empirical Jeffreys' scale used for Bayesian model comparison.}
		\label{tab:jeffreys_scale}
	\end{table}
	
	To systematically investigate the parameter space and the physical implications of the generalized LQC modifications, we categorize our Bayesian analysis into four distinct model scenarios. These scenarios are differentiated by the presence of the $\lambda$ correction term, the choice of the branch in the modified Friedmann equation~\eqref{Friedmann_equation_rho}, and whether the Barbero-Immirzi parameter $\gamma$ is fixed or treated as a free parameter. The four models are explicitly defined and labeled as follows
	\begin{itemize}
		\item \textbf{Model I (Basic LQC):} We set the correction term $\lambda = 0$ and fix the Barbero-Immirzi parameter to its standard theoretical value, $\gamma = 0.2375$. In this limit, the generalized Friedmann equation reduces to the standard, well-established effective LQC dynamics.
		
		\item \textbf{Model II (Positive Branch, Fixed $\gamma$):} We activate the $\lambda$ modification ($\lambda$ varies freely) and select the positive branch of the generalized Friedmann equation \eqref{Friedmann_equation_rho}. The Barbero-Immirzi parameter remains fixed at the theoretical prior $\gamma = 0.2375$. This scenario isolates the pure cosmological effects of the $\lambda$ term.
		
		\item \textbf{Model III (Positive Branch, Free $\gamma$):} We retain the positive branch of Eq.~\eqref{Friedmann_equation_rho} but promote the Barbero-Immirzi parameter $\gamma$ to a free parameter in the Bayesian sampling. This scenario aims to derive completely data-driven cosmological constraints on $\gamma$, rigorously testing whether the late-time observational data aligns with the theoretical prediction of LQG.
		
		\item \textbf{Model IV (Negative Branch, Free $\gamma$):} Finally, we select the negative branch of the generalized Friedmann equation Eq.~\eqref{Friedmann_equation_rho} while simultaneously treating $\gamma$ as a free parameter. This allows us to comprehensively explore the alternative dynamical trajectory permitted by the generalized theory and assess its phenomenological viability against the observational datasets.
	\end{itemize}
	
	By systematically comparing these four scenarios, we can not only evaluate the validity of the $\lambda$ modification but also ascertain the observational bounds on the fundamental parameter $\gamma$ from a purely cosmological perspective.
	
	\begin{table}[htbp]
			\centering
			\renewcommand{\arraystretch}{1.4}
			\setlength{\tabcolsep}{5pt}
			\begin{tabular}{lcccc}
				\hline\hline
				Parameter & \textbf{Model I} & \textbf{Model II} & \textbf{Model III} & \textbf{Model IV} \\
				\hline
				$\Omega_{m0}$ & $0.2989^{+0.0061}_{-0.0072}$ & $0.2983^{+0.0064}_{-0.0072}$ & $0.2994^{+0.0062}_{-0.0082}$ & $0.2988\pm 0.0097$ \\
				$\Omega_{de 0}$ & $0.7006^{+0.0074}_{-0.0062}$ & $0.7007\pm 0.0095$ & $0.7000^{+0.0081}_{-0.0067}$ & $0.7009^{+0.0071}_{-0.0060}$ \\
				$H_0$ & $68.92\pm 0.28$ & $68.91^{+0.28}_{-0.31}$ & $68.95\pm 0.29$ & $68.91\pm 0.29 $ \\
				$m_{\nu}$ [eV] & $< 0.22 $ & $< 0.213$ & $< 0.246$ & $< 0.191$ \\
				$\lambda\gamma^2$ & $0$ (fixed) & $28.95$ & $12.06$ & $10^{-121.98}$ \\
				$\gamma$ & $0.2375$ (fixed) & $0.2375$ (fixed) & $0.4082$ & $0.2802$ \\
				\hline
				$\ln \mathcal{Z}$ & $-770.8$ & $-770.9$ & $-771.2$ & $-770.3$ \\
				$\Delta \ln \mathcal{Z}$ & $0$ & $-0.1$ & $-0.4$ & $0.5$ \\
				\hline\hline
			\end{tabular}
			\caption{The marginalized mean values and $1\sigma$ ($68\%$ C.L.) uncertainties for the standard cosmological parameters across the four analyzed scenarios. For the neutrino mass $m_\nu$, the $95\%$ upper bounds are typically reported. Notably, the values provided for the LQC parameters ($\lambda\gamma^2$ and $\gamma$) represent their most probable values. The statistical performance is quantified by the log-Bayesian evidence $\ln \mathcal{Z}$, with the Bayes factor $\Delta \ln \mathcal{Z}$ computed relative to Model I.}
			\label{tab:polychord_results}
	\end{table}
	
	\begin{figure}[htbp]
		\centering
		\begin{minipage}{0.8\linewidth}
			\centering
			\includegraphics[width=\linewidth]{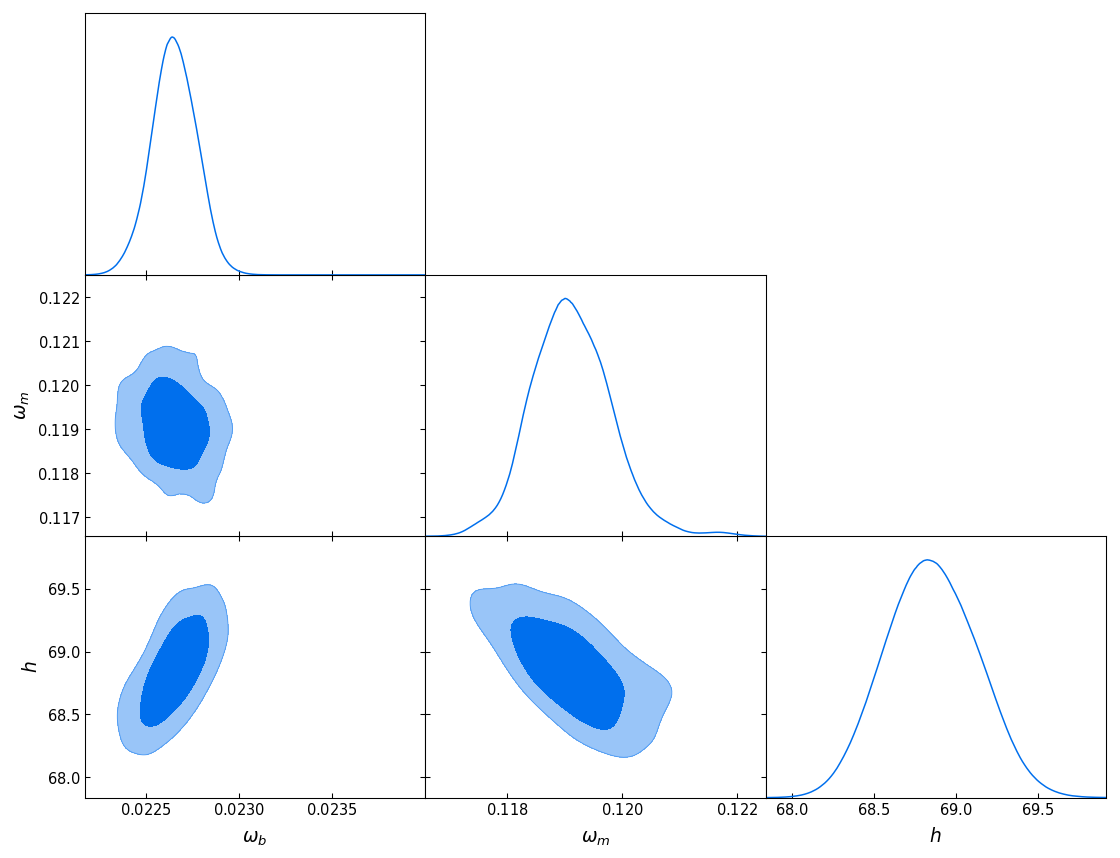}
		\end{minipage}
		\caption{One-dimensional marginalized posterior distributions and two-dimensional joint confidence contours for the standard cosmological parameters $\omega_b$, $\omega_m$, and $h$ under Model I. The inner and outer shaded regions in the 2D plots correspond to the 68\% and 95\% confidence levels, respectively.}
		\label{distribution}
	\end{figure}

	\begin{figure}[htbp]
		\centering
		\begin{minipage}{0.3\linewidth}
			\centering
			\includegraphics[width=\linewidth]{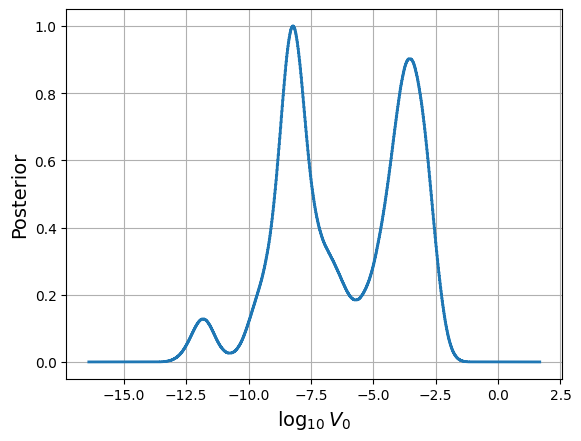}
		\end{minipage}
		\centering
		\begin{minipage}{0.3\linewidth}
			\centering
			\includegraphics[width=\linewidth]{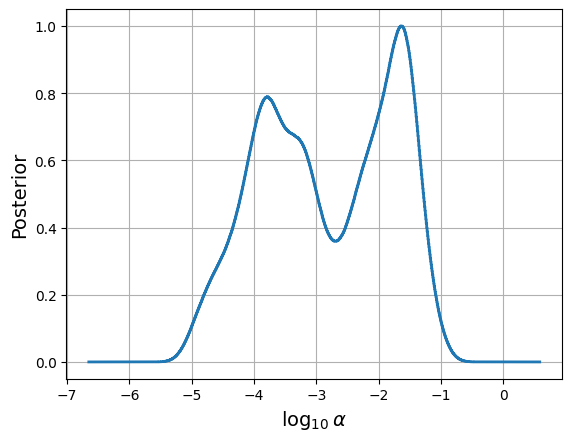}
		\end{minipage}
		\centering
		\begin{minipage}{0.3\linewidth}
			\centering
			\includegraphics[width=\linewidth]{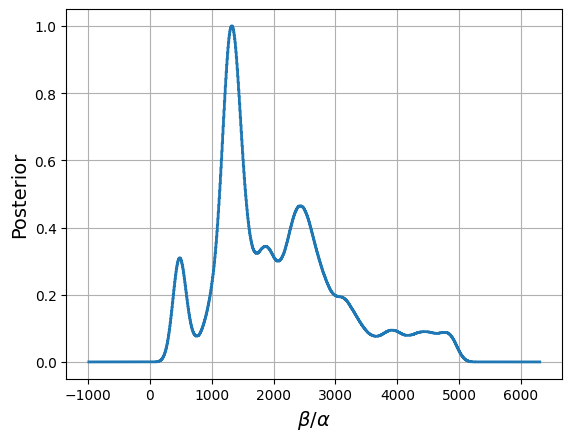}
		\end{minipage}
		\caption{One-dimensional marginalized posterior distributions for the scalar field potential parameters under Model I. (\textbf{a}) The logarithmic energy scale of the potential, $\log_{10} V_0$. (\textbf{b}) The logarithmic shape parameter, $\log_{10} \alpha$. (\textbf{c}) The parameter ratio, $\beta/\alpha$. The highly non-Gaussian, multi-modal profiles exhibited across all three distributions indicate complex parameter degeneracies and the existence of multiple observationally favored regions within the extended model.}
		\label{one dimension dis v0,alpha,beta_alpha}
	\end{figure}
	
	\begin{figure}[htbp]
		\centering
		\begin{minipage}{0.4\linewidth}
			\centering
			\includegraphics[width=\linewidth]{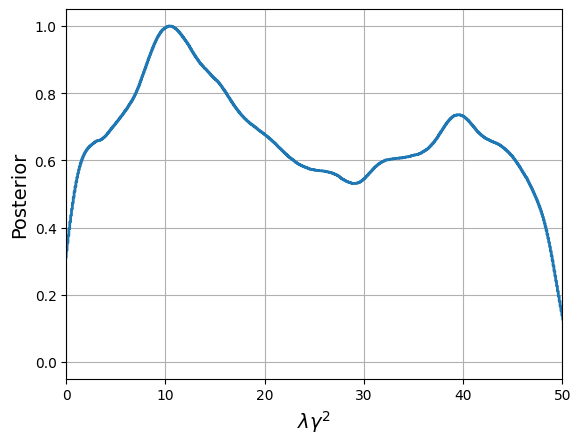}
		\end{minipage}
		\begin{minipage}{0.4\linewidth}
			\centering
			\includegraphics[width=\linewidth]{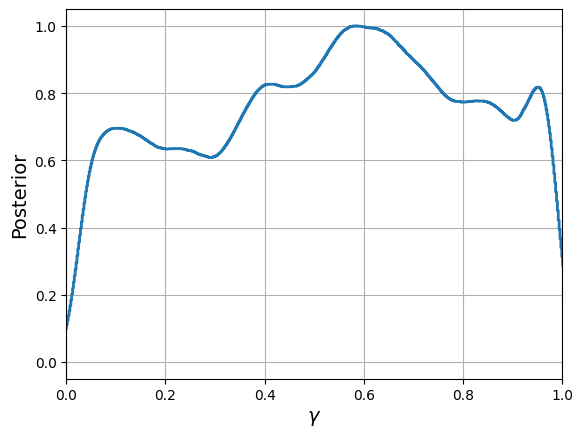}
		\end{minipage}
		\\
		\begin{minipage}{0.4\linewidth}
			\centering
			\includegraphics[width=\linewidth]{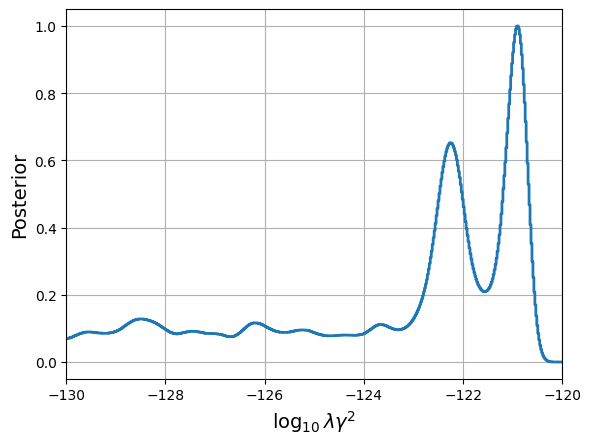}
		\end{minipage}
		\begin{minipage}{0.4\linewidth}
			\centering
			\includegraphics[width=\linewidth]{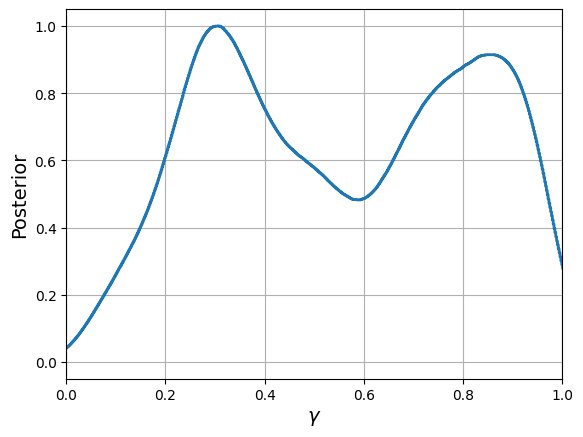}
		\end{minipage}
		\caption{One-dimensional marginalized posterior distributions for the LQC parameters. The top row corresponds to Model III, illustrating the constraints on the combined parameter $\lambda\gamma^2$ and (\textbf{b}) the Barbero-Immirzi parameter $\gamma$. The bottom row corresponds to Model IV, depicting the constraints on the logarithmic parameter $\log_{10} \lambda\gamma^2$ and (\textbf{d}) $\gamma$. The broad and multi-modal profiles suggest that these fundamental quantum geometric parameters exhibit complex parameter space topologies when constrained by current cosmological datasets.}
		\label{lg2_gamma}
	\end{figure}
	
	The numerical results of our comprehensive Bayesian analysis are summarized in Table~\ref{tab:polychord_results}, which details the marginalized mean values, uncertainties, and most probable values for the parameter space across all four models. As expected, the standard background parameters ($\Omega_{m0}$, $\Omega_{de0}$, $H_0$, and $m_\nu$) exhibit remarkable stability across all four models. Furthermore, the statistical performance of each scenario is quantified by the log-Bayesian evidence $\ln \mathcal{Z}$. As the evidence indicates, the statistical differences among the four models are remarkably small, with Model IV exhibiting only a marginal preference over the other three scenarios. 
	
	 Fig.~\ref{one dimension dis v0,alpha,beta_alpha} reveals highly non-Gaussian, multi-modal distributions for the scalar potential parameters ($\log_{10} V_0, \log_{10} \alpha, \beta/\alpha$). This multimodal profile indicates significant structural degeneracies, suggesting that multiple combinations of the potential's height and slope can equally mimic the required late-time accelerating trajectory. Similarly, the fundamental quantum parameters ($\lambda\gamma^2$ and $\gamma$) remain loosely constrained by late-time probes. As shown in Fig.~\ref{lg2_gamma}, $\gamma$ exhibits remarkably broad posteriors, failing to sharply isolate the theoretical LQG value. However, an intriguing exception emerges in Model IV. The posterior for $\gamma$ displays a distinct, dominant probability peak at a best-fit value of $\gamma = 0.2802$. Notably, this most probable value lies remarkably close to the standard theoretical expectation of $\gamma \approx 0.2375$. In Model III, the generalized correction term $\lambda\gamma^2$ also remains loosely constrained by the datasets. Its posterior distribution yields a most probable value of $\lambda\gamma^2 \approx 12.06$. Notably, this empirically favored value deviates significantly from both $\lambda\gamma^2 = 0$ and $\lambda\gamma^2 = \gamma^2$. In Model IV, the parameter combination $\lambda\gamma^2$ acts as an effective cosmological constant term. Its posterior distribution exhibits two distinct peaks, which correspond to two fundamentally different explanations for dark energy, namely a dynamical dark energy fluid and a strict cosmological constant. This bimodal profile indicates that current observational datasets are not yet sensitive enough to definitively distinguish between these two paradigms.

	\section{Conclusions}
	In this paper, we have proposed and comprehensively investigated a non-singular quintessential inflation model embedded within the effective dynamics of Loop Quantum Cosmology (LQC) based on a Generalized Regularization Scheme. By introducing a unique coupling between the quintessential scalar field and Mass-Varying Neutrinos (MaVaNs), this framework successfully unifies the primordial inflationary paradigm with late-time cosmic acceleration using a single scalar field, while naturally resolving the initial Big Bang singularity.
	
	During the extremely early epoch, quantum geometry effects successfully replace the initial singularity with a smooth quantum bounce. Immediately following the bounce, the universe enters a kinetic-dominated superinflation phase characterized by a rapidly increasing expansion rate ($\dot{H}>0$). We have demonstrated that the generalized regularization parameter $\lambda$ plays a critical role during this era by modulating the maximum expansion rate ($H_{\text{max}}$) and the corresponding intensity of the Hubble friction. Furthermore, this Planck-scale modification cascades forward to leave distinct, observable imprints on the primordial perturbation spectra, systematically shifting the scalar spectral index $n_s$ and the tensor-to-scalar ratio $r$ across the entire observable window.
	
	During the post-inflationary radiation- and matter-dominated eras, the scalar field settles into a subdominant tracking trajectory. As the universe expands and cools, neutrinos become non-relativistic, activating the coupling mechanism. This backreaction force effectively freezes the scalar field, naturally driving the universe into the observed dark energy epoch.
	
	To test the model's viability, we performed a Bayesian analysis using late-time background datasets (Planck priors, DESI BAO, and Pantheon+). In Model IV, the Barbero-Immirzi parameter $\gamma$ displays a distinct peak at $\gamma = 0.2802$—remarkably close to the theoretical loop quantum gravity expectation of $\gamma \approx 0.2375$. Furthermore, the quantum correction term $\lambda\gamma^2$ exhibits a striking bimodal profile, indicating a profound degeneracy between a dynamical dark energy fluid and a strict cosmological constant. However, relying solely on late-time background evolution data is insufficient to tightly constrain these fundamental parameters or definitively break such degeneracies. Therefore, our future work will extend beyond the background dynamics to investigate the evolution of cosmological perturbations. By confronting the model directly with full CMB power spectrum data\cite{planck2020}, we aim to uncover the unique microscopic imprints this generalized LQC framework leaves on primordial observables.
	
	\section*{Acknowledgments}\small
This work is supported by the National Natural Science Foundation of China (NSFC) with Grant No.12275087.

	\bibliographystyle{unsrt}
	\bibliography{ref.bib}

@article{Peebles_1999,
	title={Quintessential inflation},
	volume={59},
	ISSN={1089-4918},
	url={http://dx.doi.org/10.1103/PhysRevD.59.063505},
	DOI={10.1103/physrevd.59.063505},
	number={6},
	journal={Physical Review D},
	publisher={American Physical Society (APS)},
	author={Peebles, P. J. E. and Vilenkin, A.},
	year={1999},
	month=Feb }

@article{de_Haro_2021,
	title={A Review of Quintessential Inflation},
	volume={9},
	ISSN={2075-4434},
	url={http://dx.doi.org/10.3390/galaxies9040073},
	DOI={10.3390/galaxies9040073},
	number={4},
	journal={Galaxies},
	publisher={MDPI AG},
	author={de Haro, Jaume and Aresté Saló, Llibert},
	year={2021},
	month=Oct, pages={73} }

@article{Dimopoulos_2002,
	title={Modeling quintessential inflation},
	volume={18},
	ISSN={0927-6505},
	url={http://dx.doi.org/10.1016/S0927-6505(02)00115-9},
	DOI={10.1016/s0927-6505(02)00115-9},
	number={3},
	journal={Astroparticle Physics},
	publisher={Elsevier BV},
	author={Dimopoulos, K. and Valle, J.W.F.},
	year={2002},
	month=Dec, pages={287–306} }

@article{Brookfield_2006,
	title={Cosmology with Massive Neutrinos Coupled to Dark Energy},
	volume={96},
	ISSN={1079-7114},
	url={http://dx.doi.org/10.1103/PhysRevLett.96.061301},
	DOI={10.1103/physrevlett.96.061301},
	number={6},
	journal={Physical Review Letters},
	publisher={American Physical Society (APS)},
	author={Brookfield, A. W. and van de Bruck, C. and Mota, D. F. and Tocchini-Valentini, D.},
	year={2006},
	month=Feb }

@article{Ashtekar_2006,
	title={Quantum Nature of the Big Bang},
	volume={96},
	ISSN={1079-7114},
	url={http://dx.doi.org/10.1103/PhysRevLett.96.141301},
	DOI={10.1103/physrevlett.96.141301},
	number={14},
	journal={Physical Review Letters},
	publisher={American Physical Society (APS)},
	author={Ashtekar, Abhay and Pawlowski, Tomasz and Singh, Parampreet},
	year={2006},
	month=Apr }

@article{Ashtekar_2006_2,
	title={Quantum nature of the big bang: Improved dynamics},
	volume={74},
	ISSN={1550-2368},
	url={http://dx.doi.org/10.1103/PhysRevD.74.084003},
	DOI={10.1103/physrevd.74.084003},
	number={8},
	journal={Physical Review D},
	publisher={American Physical Society (APS)},
	author={Ashtekar, Abhay and Pawlowski, Tomasz and Singh, Parampreet},
	year={2006},
	month=Oct }

@misc{ashtekar2003mathematicalstructureloopquantum,
	title={Mathematical structure of loop quantum cosmology}, 
	author={Abhay Ashtekar and Martin Bojowald and Jerzy Lewandowski},
	year={2003},
	eprint={gr-qc/0304074},
	archivePrefix={arXiv},
	primaryClass={gr-qc},
	url={https://arxiv.org/abs/gr-qc/0304074}, 
}

@article{Bojowald_2001,
	title={Absence of a Singularity in Loop Quantum Cosmology},
	volume={86},
	ISSN={1079-7114},
	url={http://dx.doi.org/10.1103/PhysRevLett.86.5227},
	DOI={10.1103/physrevlett.86.5227},
	number={23},
	journal={Physical Review Letters},
	publisher={American Physical Society (APS)},
	author={Bojowald, Martin},
	year={2001},
	month=June, pages={5227–5230} }

@article{Ashtekar_2010,
	title={Loop quantum cosmology and slow roll inflation},
	volume={694},
	ISSN={0370-2693},
	url={http://dx.doi.org/10.1016/j.physletb.2010.09.058},
	DOI={10.1016/j.physletb.2010.09.058},
	number={2},
	journal={Physics Letters B},
	publisher={Elsevier BV},
	author={Ashtekar, Abhay and Sloan, David},
	year={2010},
	month=Nov, pages={108–112} }

@article{Singh_2006,
	title={Nonsingular bouncing universes in loop quantum cosmology},
	volume={74},
	ISSN={1550-2368},
	url={http://dx.doi.org/10.1103/PhysRevD.74.043510},
	DOI={10.1103/physrevd.74.043510},
	number={4},
	journal={Physical Review D},
	publisher={American Physical Society (APS)},
	author={Singh, Parampreet and Vandersloot, Kevin and Vereshchagin, G. V.},
	year={2006},
	month=Aug }

@article{Bojowald_2002,
	title={Inflation from Quantum Geometry},
	volume={89},
	ISSN={1079-7114},
	url={http://dx.doi.org/10.1103/PhysRevLett.89.261301},
	DOI={10.1103/physrevlett.89.261301},
	number={26},
	journal={Physical Review Letters},
	publisher={American Physical Society (APS)},
	author={Bojowald, Martin},
	year={2002},
	month=Dec }

@article{Copeland_2008,
	title={Superinflation in loop quantum cosmology},
	volume={77},
	ISSN={1550-2368},
	url={http://dx.doi.org/10.1103/PhysRevD.77.023510},
	DOI={10.1103/physrevd.77.023510},
	number={2},
	journal={Physical Review D},
	publisher={American Physical Society (APS)},
	author={Copeland, E. J. and Mulryne, D. J. and Nunes, N. J. and Shaeri, M.},
	year={2008},
	month=Jan }

@article{Ashtekar_2011,
	title={Probability of inflation in loop quantum cosmology},
	volume={43},
	ISSN={1572-9532},
	url={http://dx.doi.org/10.1007/s10714-011-1246-y},
	DOI={10.1007/s10714-011-1246-y},
	number={12},
	journal={General Relativity and Gravitation},
	publisher={Springer Science and Business Media LLC},
	author={Ashtekar, Abhay and Sloan, David},
	year={2011},
	month=Aug, pages={3619–3655} }

@article{Fardon_2004,
	title={Dark energy from mass varying neutrinos},
	volume={2004},
	ISSN={1475-7516},
	url={http://dx.doi.org/10.1088/1475-7516/2004/10/005},
	DOI={10.1088/1475-7516/2004/10/005},
	number={10},
	journal={Journal of Cosmology and Astroparticle Physics},
	publisher={IOP Publishing},
	author={Fardon, Rob and Nelson, Ann E and Weiner, Neal},
	year={2004},
	month=Oct, pages={005–005} }

@article{Wetterich_2007,
	title={Growing neutrinos and cosmological selection},
	volume={655},
	ISSN={0370-2693},
	url={http://dx.doi.org/10.1016/j.physletb.2007.08.060},
	DOI={10.1016/j.physletb.2007.08.060},
	number={5-6},
	journal={Physics Letters B},
	publisher={Elsevier BV},
	author={Wetterich, C.},
	year={2007},
	month=Nov, pages={201–208} }

@article{Amendola_2008,
	title={Quintessence cosmologies with a growing matter component},
	volume={78},
	ISSN={1550-2368},
	url={http://dx.doi.org/10.1103/PhysRevD.78.023015},
	DOI={10.1103/physrevd.78.023015},
	number={2},
	journal={Physical Review D},
	publisher={American Physical Society (APS)},
	author={Amendola, Luca and Baldi, Marco and Wetterich, Christof},
	year={2008},
	month=July }

@article{Afshordi_2005,
	title={Instability of dark energy with mass-varying neutrinos},
	volume={72},
	ISSN={1550-2368},
	url={http://dx.doi.org/10.1103/PhysRevD.72.065024},
	DOI={10.1103/physrevd.72.065024},
	number={6},
	journal={Physical Review D},
	publisher={American Physical Society (APS)},
	author={Afshordi, Niayesh and Zaldarriaga, Matias and Kohri, Kazunori},
	year={2005},
	month=Sept }

@article{Mota_2008,
	title={Neutrino clustering in growing neutrino quintessence},
	volume={663},
	ISSN={0370-2693},
	url={http://dx.doi.org/10.1016/j.physletb.2008.03.060},
	DOI={10.1016/j.physletb.2008.03.060},
	number={3},
	journal={Physics Letters B},
	publisher={Elsevier BV},
	author={Mota, D.F. and Pettorino, V. and Robbers, G. and Wetterich, C.},
	year={2008},
	month=May, pages={160–164} }

@article{Zhang_2021,
	title={Loop quantum gravity and cosmological constant},
	volume={823},
	ISSN={0370-2693},
	url={http://dx.doi.org/10.1016/j.physletb.2021.136770},
	DOI={10.1016/j.physletb.2021.136770},
	journal={Physics Letters B},
	publisher={Elsevier BV},
	author={Zhang, Xiangdong and Long, Gaoping and Ma, Yongge},
	year={2021},
	month=Dec, pages={136770} }

@article{Abdul_Karim_2025,
	title={DESI DR2 results. II. Measurements of baryon acoustic oscillations and cosmological constraints},
	volume={112},
	ISSN={2470-0029},
	DOI={10.1103/tr6y-kpc6},
	number={8},
	journal={Physical Review D},
	publisher={American Physical Society (APS)},
	author={Abdul Karim, M. and others},
	year={2025},
	month=Oct 
}

@article{Chen_2019,
	title={Distance priors from Planck final release},
	volume={2019},
	ISSN={1475-7516},
	url={http://dx.doi.org/10.1088/1475-7516/2019/02/028},
	DOI={10.1088/1475-7516/2019/02/028},
	number={02},
	journal={Journal of Cosmology and Astroparticle Physics},
	publisher={IOP Publishing},
	author={Chen, Lu and Huang, Qing-Guo and Wang, Ke},
	year={2019},
	month=Feb, pages={028–028} }

@article{Brout_2022,
	title={The Pantheon+ Analysis: Cosmological Constraints},
	volume={938},
	ISSN={1538-4357},
	url={http://dx.doi.org/10.3847/1538-4357/ac8e04},
	DOI={10.3847/1538-4357/ac8e04},
	number={2},
	journal={The Astrophysical Journal},
	publisher={American Astronomical Society},
	author={Brout, Dillon and others},
	year={2022},
	month=Oct, pages={110} }

@article{Yang_2009,
	title={Alternative quantization of the Hamiltonian in loop quantum cosmology},
	volume={682},
	ISSN={0370-2693},
	url={http://dx.doi.org/10.1016/j.physletb.2009.10.072},
	DOI={10.1016/j.physletb.2009.10.072},
	number={1},
	journal={Physics Letters B},
	publisher={Elsevier BV},
	author={Yang, Jinsong and Ding, You and Ma, Yongge},
	year={2009},
	month=Nov, pages={1–7} }

@article{HAN_2007,
	title={FUNDAMENTAL STRUCTURE OF LOOP QUANTUM GRAVITY},
	volume={16},
	ISSN={1793-6594},
	url={http://dx.doi.org/10.1142/S0218271807010894},
	DOI={10.1142/s0218271807010894},
	number={09},
	journal={International Journal of Modern Physics D},
	publisher={World Scientific Pub Co Pte Lt},
	author={HAN, MUXIN and MA, YONGGE and HUANG, WEIMING},
	year={2007},
	month=Sept, pages={1397–1474} }

@article{Ashtekar_2004,
	title={Background independent quantum gravity: a status report},
	volume={21},
	ISSN={1361-6382},
	url={http://dx.doi.org/10.1088/0264-9381/21/15/R01},
	DOI={10.1088/0264-9381/21/15/r01},
	number={15},
	journal={Classical and Quantum Gravity},
	publisher={IOP Publishing},
	author={Ashtekar, Abhay and Lewandowski, Jerzy},
	year={2004},
	month=July, pages={R53–R152} }

@article{Geng_2017,
	title={Observational constraints on successful model of quintessential Inflation},
	volume={2017},
	ISSN={1475-7516},
	url={http://dx.doi.org/10.1088/1475-7516/2017/06/011},
	DOI={10.1088/1475-7516/2017/06/011},
	number={06},
	journal={Journal of Cosmology and Astroparticle Physics},
	publisher={IOP Publishing},
	author={Geng, Chao-Qiang and Lee, Chung-Chi and Sami, M. and Saridakis, Emmanuel N. and Starobinsky, Alexei A.},
	year={2017},
	month=June, pages={011–011} }

@article{Geng_2015,
	title={Quintessential inflation with canonical and noncanonical scalar fields and Planck 2015 results},
	volume={92},
	ISSN={1550-2368},
	url={http://dx.doi.org/10.1103/PhysRevD.92.023522},
	DOI={10.1103/physrevd.92.023522},
	number={2},
	journal={Physical Review D},
	publisher={American Physical Society (APS)},
	author={Geng, Chao-Qiang and Hossain, Md. Wali and Myrzakulov, R. and Sami, M. and Saridakis, Emmanuel N.},
	year={2015},
	month=July }

@article{Bojowald_2011,
	title={Observational test of inflation in loop quantum cosmology},
	volume={2011},
	ISSN={1475-7516},
	url={http://dx.doi.org/10.1088/1475-7516/2011/11/046},
	DOI={10.1088/1475-7516/2011/11/046},
	number={11},
	journal={Journal of Cosmology and Astroparticle Physics},
	publisher={IOP Publishing},
	author={Bojowald, Martin and Calcagni, Gianluca and Tsujikawa, Shinji},
	year={2011},
	month=Nov, pages={046–046} }

@article{Agullo_2012,
	title={Quantum Gravity Extension of the Inflationary Scenario},
	volume={109},
	ISSN={1079-7114},
	url={http://dx.doi.org/10.1103/PhysRevLett.109.251301},
	DOI={10.1103/physrevlett.109.251301},
	number={25},
	journal={Physical Review Letters},
	publisher={American Physical Society (APS)},
	author={Agullo, Ivan and Ashtekar, Abhay and Nelson, William},
	year={2012},
	month=Dec }

@article{Artymowski_2009,
	title={Loop Quantum Cosmology: holonomy corrections to inflationary models},
	volume={2009},
	ISSN={1475-7516},
	url={http://dx.doi.org/10.1088/1475-7516/2009/01/004},
	DOI={10.1088/1475-7516/2009/01/004},
	number={01},
	journal={Journal of Cosmology and Astroparticle Physics},
	publisher={IOP Publishing},
	author={Artymowski, Michał and Lalak, Zygmunt and Szulc, Łukasz},
	year={2009},
	month=Jan, pages={004–004} }

@article{Barboza_2022,
	title={Constraining the Barbero-Immirzi parameter from the duration of inflation in loop quantum cosmology},
	volume={106},
	ISSN={2470-0029},
	url={http://dx.doi.org/10.1103/PhysRevD.106.103535},
	DOI={10.1103/physrevd.106.103535},
	number={10},
	journal={Physical Review D},
	publisher={American Physical Society (APS)},
	author={Barboza, L. N. and Levy, G. L. L. W. and Graef, L. L. and O. Ramos  Rudnei},
	year={2022},
	month=Nov }

@article{Bhardwaj_2019,
	title={Inflation in loop quantum cosmology},
	volume={99},
	ISSN={2470-0029},
	url={http://dx.doi.org/10.1103/PhysRevD.99.063520},
	DOI={10.1103/physrevd.99.063520},
	number={6},
	journal={Physical Review D},
	publisher={American Physical Society (APS)},
	author={Bhardwaj, Anshuman and Copeland, Edmund J. and Louko, Jorma},
	year={2019},
	month=Mar }

@article{Bonga_2016,
	title={Inflation with the Starobinsky potential in loop quantum cosmology},
	volume={48},
	ISSN={1572-9532},
	url={http://dx.doi.org/10.1007/s10714-016-2071-0},
	DOI={10.1007/s10714-016-2071-0},
	number={6},
	journal={General Relativity and Gravitation},
	publisher={Springer Science and Business Media LLC},
	author={Bonga, Béatrice and Gupt, Brajesh},
	year={2016},
	month=May }

@article{Zhang_2007,
	title={Inflationary universe in loop quantum cosmology},
	volume={2007},
	ISSN={1475-7516},
	url={http://dx.doi.org/10.1088/1475-7516/2007/08/012},
	DOI={10.1088/1475-7516/2007/08/012},
	number={08},
	journal={Journal of Cosmology and Astroparticle Physics},
	publisher={IOP Publishing},
	author={Zhang, Xin and Ling, Yi},
	year={2007},
	month=Aug, pages={012–012} }

@article{Bonga_2016_2,
	title={Phenomenological investigation of a quantum gravity extension of inflation with the Starobinsky potential},
	volume={93},
	ISSN={2470-0029},
	url={http://dx.doi.org/10.1103/PhysRevD.93.063513},
	DOI={10.1103/physrevd.93.063513},
	number={6},
	journal={Physical Review D},
	publisher={American Physical Society (APS)},
	author={Bonga, Béatrice and Gupt, Brajesh},
	year={2016},
	month=Mar }

@article{planck2020,
	title={Planck 2018 results: X. Constraints on inflation},
	volume={641},
	ISSN={1432-0746},
	url={http://dx.doi.org/10.1051/0004-6361/201833887},
	DOI={10.1051/0004-6361/201833887},
	journal={Astronomy Astrophysics},
	publisher={EDP Sciences},
	author={ Akrami, Y. and others},
	year={2020},
	month=Sept, pages={A10} }

@article{Liddle_2003,
	title={How long before the end of inflation were observable perturbations produced?},
	volume={68},
	ISSN={1089-4918},
	url={http://dx.doi.org/10.1103/PhysRevD.68.103503},
	DOI={10.1103/physrevd.68.103503},
	number={10},
	journal={Physical Review D},
	publisher={American Physical Society (APS)},
	author={Liddle, Andrew R. and Leach, Samuel M.},
	year={2003},
	month=Nov }

@article{Dimopoulos_2017,
	title={Quintessential inflation with {$\alpha$}-attractors},
	volume={2017},
	ISSN={1475-7516},
	url={http://dx.doi.org/10.1088/1475-7516/2017/06/027},
	DOI={10.1088/1475-7516/2017/06/027},
	number={06},
	journal={Journal of Cosmology and Astroparticle Physics},
	publisher={IOP Publishing},
	author={Dimopoulos, Konstantinos and Owen, Charlotte},
	year={2017},
	month=June, pages={027–027} }

@article{PhysRevD.35.2955,
	title = {Gravitational particle creation and inflation},
	author = {Ford, L. H.},
	journal = {Phys. Rev. D},
	volume = {35},
	issue = {10},
	pages = {2955--2960},
	numpages = {0},
	year = {1987},
	month = {May},
	publisher = {American Physical Society},
	doi = {10.1103/PhysRevD.35.2955},
	url = {https://link.aps.org/doi/10.1103/PhysRevD.35.2955}
}

@article{Felder_1999,
	title={Instant preheating},
	volume={59},
	ISSN={1089-4918},
	url={http://dx.doi.org/10.1103/PhysRevD.59.123523},
	DOI={10.1103/physrevd.59.123523},
	number={12},
	journal={Physical Review D},
	publisher={American Physical Society (APS)},
	author={Felder, Gary and Kofman, Lev and Linde, Andrei},
	year={1999},
	month=May }

@article{PhysRevD.42.453,
	title = {Energy density of relic gravity waves from inflation},
	author = {Sahni, Varun},
	journal = {Phys. Rev. D},
	volume = {42},
	issue = {2},
	pages = {453--463},
	numpages = {0},
	year = {1990},
	month = {Jul},
	publisher = {American Physical Society},
	doi = {10.1103/PhysRevD.42.453},
	url = {https://link.aps.org/doi/10.1103/PhysRevD.42.453}
}

@article{Torrado_2021,
	title={Cobaya: code for Bayesian analysis of hierarchical physical models},
	volume={2021},
	ISSN={1475-7516},
	url={http://dx.doi.org/10.1088/1475-7516/2021/05/057},
	DOI={10.1088/1475-7516/2021/05/057},
	number={05},
	journal={Journal of Cosmology and Astroparticle Physics},
	publisher={IOP Publishing},
	author={Torrado, Jesús and Lewis, Antony},
	year={2021},
	month=May, pages={057} }

@article{Handley_2015,
	title={<scp>polychord</scp>: nested sampling for cosmology},
	volume={450},
	ISSN={1745-3925},
	url={http://dx.doi.org/10.1093/mnrasl/slv047},
	DOI={10.1093/mnrasl/slv047},
	number={1},
	journal={Monthly Notices of the Royal Astronomical Society: Letters},
	publisher={Oxford University Press (OUP)},
	author={Handley, W. J. and Hobson, M. P. and Lasenby, A. N.},
	year={2015},
	month=Apr, pages={L61–L65} }

@article{Handley_2015_2,
	title={polychord: next-generation nested sampling},
	volume={453},
	ISSN={1365-2966},
	url={http://dx.doi.org/10.1093/mnras/stv1911},
	DOI={10.1093/mnras/stv1911},
	number={4},
	journal={Monthly Notices of the Royal Astronomical Society},
	publisher={Oxford University Press (OUP)},
	author={Handley, W. J. and Hobson, M. P. and Lasenby, A. N.},
	year={2015},
	month=Sept, pages={4385–4399} }

@article{Kass01061995,
	author = {Robert E. Kass and Adrian E. Raftery},
	title = {Bayes Factors},
	journal = {Journal of the American Statistical Association},
	volume = {90},
	number = {430},
	pages = {773--795},
	year = {1995},
	publisher = {Taylor \& Francis},
	doi = {10.1080/01621459.1995.10476572},
	URL = { 
	https://doi.org/10.1080/01621459.1995.10476572
	},
	eprint = {
	https://doi.org/10.1080/01621459.1995.10476572}
}

@article{Trotta_2008,
	title={Bayes in the sky: Bayesian inference and model selection in cosmology},
	volume={49},
	ISSN={1366-5812},
	url={http://dx.doi.org/10.1080/00107510802066753},
	DOI={10.1080/00107510802066753},
	number={2},
	journal={Contemporary Physics},
	publisher={Informa UK Limited},
	author={Trotta, Roberto},
	year={2008},
	month=Mar, pages={71–104} }

@article{Riess_2022,
	title={A Comprehensive Measurement of the Local Value of the Hubble Constant with 1 km s$^{-1}$ Mpc$^{-1}$ Uncertainty from the Hubble Space Telescope and the {SH0ES} Team},
	volume={934},
	ISSN={2041-8213},
	url={http://dx.doi.org/10.3847/2041-8213/ac5c5b},
	DOI={10.3847/2041-8213/ac5c5b},
	number={1},
	journal={The Astrophysical Journal Letters},
	publisher={American Astronomical Society},
	author={Riess, Adam G. and others},
	year={2022},
	month=jul, 
	pages={L7} 
}
\end{document}